\documentclass{llncs}

\usepackage{epsfig} 

\usepackage{amsmath}
\usepackage{amssymb}
\usepackage{pstricks,pst-tree}

\usepackage{bussproofs}

\newcommand{\qm}[1]{``#1''}
\newcommand{\gramor}{\ \,|\ \,}
\newcommand{\paral}{\;|\;}
\newcommand{\labarr}[1]{\overset{#1}{\longrightarrow}}
\newcommand{\smallsum}[1]{\textstyle{\sum_{#1}\:}}
\newcommand{\smallsumb}[2]{\textstyle{\sum_{#1}^{#2}\:}}
\newcommand{\smallprodb}[2]{\textstyle{\prod_{#1}^{#2}\:}}
\newcommand{\bigfrac}[2]{\frac{\displaystyle #1}{\displaystyle #2}}
\newcommand{\defeq}{\overset{\scriptscriptstyle\Delta}{=}}
\newcommand{\ang}[1]{\langle#1\rangle}
\newcommand{\outp}[1]{\overline{#1}}

\newcommand{\cl}{\hspace{-2pt}:\hspace{-2pt}}
\newcommand{\tmust}{\textrm{\textbf{must}}}
\newcommand{\tmay}{\textrm{\textbf{may}}}
\newcommand{\may}{\sqsubseteq_\tmay}
\newcommand{\must}{\sqsubseteq_\tmust}
\newcommand{\mayeq}{\approx_\tmay}
\newcommand{\musteq}{\approx_\tmust}
\newcommand{\weaker}{\sqsubseteq}
\newcommand{\eqeq}{\,\textrm{\texttt{==}}\,}

\newcommand{\ccss}{CCS$_\sigma$}
\newcommand{\ccsp}{CCS$_p$}
\newcommand{\cald}{\mathcal{D}}
\newcommand{\pom}{p_\omega}

\begin{document}

\title{Making Random Choices Invisible to the Scheduler\thanks{This
    work has been partially supported by the INRIA DREI
    \'Equipe Associ\'ee PRINTEMPS and by the INRIA ARC project
    ProNoBiS.}
}
\author{Konstantinos Chatzikokolakis \qquad Catuscia
  Palamidessi\\
  {\small INRIA and LIX, \'Ecole Polytechnique, Palaiseau, France
  \tt{\{kostas,catuscia\}@lix.polytechnique.fr}}
}
\institute{}

\thispagestyle{plain} % remove this command for the camera-ready version
\pagestyle{plain} % remove this command for the camera-ready version

\maketitle

\begin{abstract}
When dealing with process calculi and automata which express both
nondeterministic and probabilistic behavior, it is customary to
introduce the notion of scheduler to solve the nondeterminism. It has
been observed that for certain applications, notably those in
security, the scheduler needs to be restricted so not to reveal the
outcome of the protocol's random choices, or otherwise the model of
adversary would be too strong even for ``obviously correct''
protocols. We propose a process-algebraic framework in which the
control on the scheduler can be specified in syntactic terms, and we
show how to apply it to solve the problem mentioned above. We also
consider the definition of (probabilistic) may and must preorders, and
we show that they are precongruences with respect to the restricted
schedulers. Furthermore, we show that all the operators of the
language, except replication, distribute over probabilistic summation,
which is a useful property for verification.
% Finally, we discuss the relation with the ``semantic'' notion
% of scheduler which is more similar to those considered in literature. 
\end{abstract}

\section{Introduction}
\label{sec:introduction}

Security protocols, in particular those for for anonymity and fair
exchange, often use randomization to achieve their targets. Since they
usually involve more than one agent, they also give rise to concurrent
and interactive activities that can be best modeled by
nondeterminism. Thus it is convenient to specify them using a
formalism which is able to represent both \emph{probabilistic} and
\emph{nondeterministic} behavior.  Formalisms of this kind have been
explored in both Automata Theory
\cite{Vardi:85:FOCS,Hansson:89:SRTS,Yi:92:IFIP,Segala:95:PhD,Segala:95:NJC}
and in Process Algebra
\cite{Hansson:90:SRTS,Bandini:01:ICALP,Andova:02:PhD,Mislove:03:EXPRESS,Palamidessi:05:TCS,Deng:05:BookJW}.
See also \cite{Sokolova:04:BookChapt,Jonsson:2001:HANDBOOK} for
comparative and more inclusive overviews.

Due to the presence of nondeterminism, in such formalisms it is not
possible to define the probability of events in \emph{absolute}
terms. We need first to decide how each nondeterministic choice during
the execution will be solved.  This decision function is called
\emph{scheduler}. Once the scheduler is fixed, the behavior of the
system (\emph{relatively} to the given scheduler) becomes fully
probabilistic and a probability measure can be defined following
standard techniques.

It has been observed by several researchers that in security the
notion of scheduler needs to be restricted, or otherwise any secret
choice of the protocol could be revealed by making the choice of the
scheduler depend on it.  This issue was for instance one of the main
topics of discussion at the panel of CSFW 2006. We illustrate it here
with an example on anonymity.  We use the standard CCS notation, plus
a construct of probabilistic choice $P+_pQ$ representing a process
that evolves into $P$ with probability $p$ and into $Q$ with
probability $1-p$.

The following system \emph{Sys} consists of one receiver $R$ and two
senders $S, T$ which communicate via private channels $a,b$
respectively. Which of the two senders is successful is decided
probabilistically by $R$.  After reception, $R$ sends a signal
\emph{ok}.
\[
R\,\defeq\,a.\overline{\it ok}.0 \; +_{0.5} \; b.\overline{\it ok}.0 \;\;\;\;\;\; S\,\defeq\,\bar{a}.0  \;\;\;\;\;\; T\,\defeq\,\bar{b}.0 \;\;\;\;\;\; 
\it{Sys}\,\defeq\,(\nu a)(\nu b)(R \paral S \paral T)
\]
The signal {\it ok} is not private, but since it is the same in both
cases, in principle an external observer should not be able to infer
from it the identity of the sender ($S$ or $T$). So the system should
be anonymous.  However, consider a team of two attackers $A$ and $B$
defined as
\[
A\,\defeq\,{\it ok}.\bar s.0  \;\;\;\;\;\; B\,\defeq\,{\it ok}.\bar t.0
\]
and consider the parallel composition $\it{Sys} \paral A \paral B$. We
have that, under certain schedulers, the system is no longer
anonymous.  More precisely, a scheduler could leak the identity of the
sender via the channels $s, t$ by forcing $R$ to synchronize with $A$
on {\it ok} if $R$ has chosen the first alternative, and with $B$
otherwise.  This is because in general a scheduler can see the whole
history of the computation, in particular the random choices, even
those which are supposed to be private. Note that the visibility of
the synchronization channels to the scheduler is not crucial for this
example: we would have the same problem, for instance, if $S$, $T$
were both defined as $\bar{a}.0$, $R$ as $a.\overline{\it ok}.0$, and
$\it{Sys}$ as $(\nu a)((R\; +_{0.5} \; S) \paral T)$.

The above example demonstrates that, with the standard
definition of scheduler, it is not possible to represent a truly
private random choice (or a truly private nondeterministic choice,
for the matter) with the current probabilistic process calculi. This
is a clear shortcoming when we want to use these formalisms for 
the specification and verification of security protocols. 

%In such protocols the outcome of a random choice, such as a coin toss, 
%is usually meant to remain private unless
%the process communicates it to the environment. As a consequence,
%the process should be able to toss the coin even before the
%execution of the protocol and keep the outcome private until it
%is actually used during the execution. However this not possible
%if the choice is visible to the scheduler, as expressed in the
%following example (we use here CCS with value passing for brevity):
There is another issue related to verification: a private
choice has certain algebraic properties that would be useful 
in proving equivalences between processes. In fact, if the outcome 
of a choice remains private, then it should not matter at which
point of the execution the process makes such choice, until it actually uses it.
Consider for instance $A$ and $B$ defined as follows 
\begin{center}
\begin{tabular}{c@{\hspace{50pt}}c}
	$
	\begin{array}{r@{}l}
		A \defeq a(x).(
			& [x = 0] \outp{\it ok}\\
			& +_{0.5} \\
			& [x = 1] \outp{\it ok})
	\end{array}
	$ &
	$
	\begin{array}{r@{}l}
		B \defeq
			& a(x).[x = 0] \outp{\it ok}\\
			& +_{0.5} \\
			& a(x).[x = 1] \outp{\it ok} \\
	\end{array}
	$
\end{tabular}
\end{center}
Process $A$ receives a value and then decides randomly whether it will accept
the value $0$ or $1$. Process $B$ does exactly the same thing except that
the choice is performed before the reception of the value.
If the random choices in $A$ and $B$ are private, intuitively we
should have that $A$ and $B$ are equivalent ($A \approx B$). This is
because it should not matter whether the choice is done
before or after receiving a message, as long as the outcome of the
choice is completely invisible to any other process or observer.
However, consider the parallel context $C = \outp{a} 0 \paral
\outp{a} 1$.  Under any scheduler $A$ has probability at most $1/2$ to
perform $\outp{\it ok}$. With $B$, on the other hand, the scheduler can choose
between $\outp{a} 0$ and $\outp{a} 1$ based on the outcome of the
probabilistic choice, thus making the maximum probability of $\outp{\it ok}$
equal to $1$. The execution trees of $A\paral C$ and $B\paral C$ are
shown in Figure \ref{fig:exectrees}.
\begin{figure}[tb]
	\centering
	\pstree[treesep=2pt,levelsep=110pt,nodesep=5pt,treemode=R,edge=\nccurve,angleB=180,ncurvA=0,ncurvB=0]
		   {\Tr{$A\paral\bar{a} 0\paral\bar{a} 1$}}{
		\pstree[levelsep=95pt]{
			\Tr{$  ([0 = 0] \outp{ok} +_{0.5} [0 = 1]
                          \outp{ok}) \paral \bar{a} 1  $}
		}{
			\Tr{\makebox[10pt]{$\outp{ok}$}}	\ncput{\rnode{a1}{}}
			\Tr{\makebox[10pt]{$0$}}			\ncput{\rnode{b1}{}}
		}
		\pstree[levelsep=95pt]{
			\Tr{$ ([1 = 0] \outp{ok} +_{0.5} [1 = 1]
                          \outp{ok}) \paral \bar{a} 0  $}
		}{
			\Tr{\makebox[10pt]{$0$}}			\ncput{\rnode{a2}{}}
			\Tr{\makebox[10pt]{$\outp{ok}$}}	\ncput{\rnode{b2}{}}
		}
	}
	\ncarc[arcangle=30]{a1}{b1}
	\ncarc[arcangle=30]{a2}{b2}
	\\[5pt]
	\pstree[treesep=2pt,levelsep=100pt,nodesep=5pt,treemode=R,edge=\nccurve,angleB=180,ncurvA=0,ncurvB=0]
		   {\Tr{$B\paral\bar{a} 0\paral\bar{a} 1$}}{
		\pstree[levelsep=80pt]{
			\Tr{$  a(x).[x = 0] \outp{ok} \paral \bar{a}
                          {0} \paral \bar{a} 1  $}
			\ncput{\rnode{a3}{}}
		}{
			\Tr{\makebox[10pt]{$\outp{ok}$}}
			\Tr{\makebox[10pt]{$0$}}
		}
		\pstree[levelsep=80pt]{
			\Tr{$  a(x).[x = 1] \outp{ok} \paral \bar{a}
                          {0} \paral \bar{a} 1  $}
			\ncput{\rnode{b3}{}}
		}{
			\Tr{\makebox[10pt]{$0$}}
			\Tr{\makebox[10pt]{$\outp{ok}$}}
		}
	}
	\ncarc[arcangle=30]{a3}{b3}
	\caption{Execution trees for $A\paral C$ and $B \paral C$}
	\label{fig:exectrees}
\end{figure}

In general when $+_p$ represents a private choice we would like to have
\begin{equation}
	C[P +_p Q] \approx C[\tau.P] +_p C[\tau.Q]
	\label{eq:equivcontext}
\end{equation}
for all processes $P,Q$ and all contexts $C$ \emph{not containing
replication (or recursion)}. In the case of replication the above
cannot hold since $!(P +_p Q)$ makes available each time the choice
between $P$ and $Q$, while $(!\tau.P)+_p(!\tau.Q)$ chooses once and for all
which of the two ($P$ or $Q$) should be replicated. Similarly for
recursion.
The reason why we need a $\tau$ is explained in Section \ref{sec:testing}.
%prefix in front of $P$ and $Q$, in the
%right hand side of (\ref{eq:equivcontext}), is to keep $+_p$ 
%private (in case $C$ contains $+$ at the top-level).

The algebraic property (\ref{eq:equivcontext}) expresses in an abstract way the
privacy of the probabilistic choice. Moreover, this property is also useful
for the verification of security properties. The interested reader can find in
\cite{Chatzikokolakis:05:TCS} an example 
of application to a fair 
exchange protocol. In principle 
(\ref{eq:equivcontext}) should be useful for any kind of verification 
in the process algebra style. 

We propose a process-algebraic approach to the problem of hiding 
the outcome of random choices. 
Our framework is based on a  calculus obtained by adding to CCS  
an internal probabilistic choice construct\footnote{We
actually consider a variant of CCS where recursion is replaced by
replication. The two languages are not equivalent, but 
we believe that the issues regarding the differences between 
replication and recursion are orthogonal to the topics investigated in this paper.}. 
 This calculus, 
to which we refer as \ccsp{}, is a variant of the 
one studied in \cite{Deng:05:BookJW}, the main differences being that
we use replication instead than recursion, and we lift some
restrictions that were imposed in \cite{Deng:05:BookJW} to
obtain a complete axiomatization. The semantics
of \ccsp{} is given in terms of Segala's  \emph{simple probabilistic
automata} \cite{Segala:95:PhD,Bandini:01:ICALP}.

In order to limit the power of the scheduler, we extend \ccsp{} with
terms representing explicitly the notion of scheduler. The
latter interact with the original processes via a labeling system. 
This will allow to specify at the syntactic level (by a suitable
labeling) which choices should be visible to schedulers, 
and which ones should not. 

\subsection{Contribution}
The main contributions of this paper are: 
\begin{itemize}
\item A process calculus \ccss{} in which the scheduler is represented as a
  process, and whose power can therefore be controlled at the
  syntactic level. 
\item An  application of \ccss{} to 
  an extended anonymity example (the Dining
  Cryptographers Protocol, DCP). We also
  briefly outline how to extend \ccss{} so to  allow 
  the definition of private nondeterministic choice, and we apply it 
  to the DCP with nondeterministic master. To our knowledge
  this is the first formal treatment of the scheduling problem in DCP 
  and the first formalization of a nondeterministic master for the
  (probabilistic) DCP. 
\item The adaptation of the standard notions of probabilistic testing preorders
  to \ccss{}, and the ``sanity check'' that they are still
  precongruences with respect to all the operators except the
  nondeterministic sum. For the latter we have the problem
  that $P$ and $\tau.P$ are must equivalent, but  $Q + P$ and $Q + \tau.P$
  are not. This is typical for the CCS $+$: usually it does not
  preserve weak equivalences.
\item The proof that, under suitable conditions on the labelings of
  $C$, $\tau.P$ and $\tau.Q$, \ccss{} satisfies the property
  expressed by (\ref{eq:equivcontext}), where $\approx$ is probabilistic
  testing equivalence.
\end{itemize}

\subsection{Related work}

The works that are most closely related to ours are
\cite{Canetti:06:WODES,Canetti:06:DISC}. In those paper the authors
consider probabilistic automata and introduce a restriction on the
scheduler to the purpose of making them suitable to applications in
security protocols. Their approach is based on dividing the actions 
of each component of the system in
equivalence classes (\emph{tasks}). The order of execution of
different tasks is decided in advance by a so-called \emph{task
scheduler}. The remaining nondeterminism within a task is solved
by a second scheduler, which models the standard \emph{adversarial 
scheduler} of the cryptographic community. This second entity 
has limited knowledge about the other components: it sees only
the information that they communicate during execution. 

In contrast to the above approach, our definition of scheduler is
based on a labeling system, and the same action can receive different
labels during the execution, so our ``equivalence classes''
(schedulable actions with the same label) can change
dynamically. However we don't know at the moment whether this 
difference determines a separation in the expressive power. 
The main difference, anyway, is that our framework is
process-algebraic and we focus on testing preorders, their
congruence properties, and the conditions under which certain equivalences
hold. 

Another work along these lines is \cite{deAlfaro:01:CONCUR}, 
which uses partitions on the state-space to obtain partial-information
schedulers. However in that paper the authors consider a synchronous
parallel composition, so the setting is rather different. 

\subsection{Plan of the paper}
In the next section we briefly recall some basic notions. In Section
\ref{sec:ccss} we define a preliminary version of the 
language \ccss{} and of the corresponding notion of scheduler. 
In Section \ref{sec:expressiveness} we compare our
notion of scheduler with the more standard ``semantic'' notion, and 
we improve the definition of \ccss{} so to retrieve the full
expressive power of the semantic schedulers. In
Section  \ref{sec:testing} we study the probabilistic testing
preorders, their compositionality properties, and the conditions under
which  (\ref{eq:equivcontext}) holds. Section \ref{sec:application}
presents an application to security. Section \ref{sec:conclusion}
concludes. 

\section{Preliminaries}
\label{sec:preliminaries}

In this section we briefly recall some preliminary notions about
the simple probabilistic automata and \ccsp{}.

\subsection{Simple probabilistic automata \cite{Segala:95:PhD,Bandini:01:ICALP}}
A \emph{discrete probability measure} over a set $X$ is a function
$\mu:2^X \mapsto [0,1]$ such that $\mu(X) = 1$ and
$\mu(\cup_i X_i) = \sum_i\mu(X_i)$ where ${X_i}$ is a countable
family of pairwise disjoint subsets of $X$. The set of all discrete probability
measures over $X$ will be denoted by $Disc(X)$. We will denote by
$\delta(x), x\in X$ (called the \emph{Dirac measure} on $x$)
the probability measure that assigns probability $1$
to $\{x\}$. We will also denote by $\sum_i[p_i]\mu_i$ the probability
measure obtained as a convex sum of the measures $\mu_i$.

A \emph{simple probabilistic automaton}\footnote{For simplicity in the
following we will refer to a simple probabilistic automaton as 
\emph{probabilistic automaton}. Note however
that simple probabilistic automata are a subset of the probabilistic automata
defined in \cite{Segala:95:PhD,Segala:95:NJC}.} is a tuple $(S, q, A, \cald)$
where $S$ is a set of states, $q \in S$ is the \emph{initial state},
$A$ is a set of actions and $\cald \subseteq S \times A \times Disc(S)$
is a \emph{transition relation}. Intuitively, if $(s,a,\mu) \in \cald$
then there is a transition from the state $s$ performing
the action $a$ and leading to a distribution $\mu$ over the states
of the automaton. The idea is that the choice of transition
among the available ones in $\cald$ is performed nondeterministically,
and the choice of the target state among the ones allowed by $\mu$
(i.e. those states $q$ such that $\mu(q) > 0$) is performed probabilistically.

A probabilistic automaton $M$ is \emph{fully probabilistic} if from each
state of $M$ there is at most one transition available. An execution $\alpha$ of
a probabilistic automaton is a (possibly infinite) sequence $s_0 a_1 s_1 a_2 s_2 \ldots$
of alternating states and actions, 
such that $q= s_0$, and for each $i$ $(s_i,a_{i+1},\mu_i) \in \cald$  and
$\mu_i(s_{i+1}) > 0$ hold. We will use $lstate(\alpha)$ to denote the last state of a finite
execution $\alpha$, and $exec^*(M)$ and $exec(M)$ to represent the set
of all the finite and of all the executions of $M$, respectively.

A \emph{scheduler} of a probabilistic automaton $M = (S,q,A,\cald)$
is a function
\[
	\zeta: exec^*(M) \mapsto \cald
\]
such that $\zeta(\alpha) = (s, a, \mu) \in \cald$ implies that $s = lstate(\alpha)$.
The idea is that a scheduler selects a transition
among the ones available in $\cald$ and it can base his decision on the
history of the execution. The \emph{execution tree} of $M$ relative to the
scheduler $\zeta$, denoted by $etree(M,\zeta)$, is a fully probabilistic
automaton $M' = (S',q',A',\cald')$ such that $S' \subseteq exec(M)$,
$q' = q$, $A' = A$, and $(\alpha,a,\mu') \in \cald'$ if and only if
$\zeta(\alpha) = (lstate(\alpha), a, \mu)$ for some $\mu$ and
$\mu'(\alpha a s) = \mu(s)$. Intuitively, $etree(M,\zeta)$ is produced by
unfolding the executions of $M$ and resolving all deterministic choices
using $\zeta$. Note that $etree(M,\zeta)$ is a simple\footnote{This 
is true because we do not consider probabilistic schedulers.
If we considered such schedulers then the execution tree would no
longer be a simple automaton.} and fully probabilistic
automaton.

\subsection{CCS with internal probabilistic choice}
\label{sec:ccsp}
Let $a$ range over a countable set of \emph{channel names}. 
%\ccsp{} is an extension of the classical CCS with internal probabilistic choice.
The syntax of \ccsp{} is the following:
\[
\begin{array}[t]{@{\textrm{\hspace{20pt}}}l@{}l@{\textrm{\hspace{20pt}}}l}
	\multicolumn{2}{l}{\alpha ::= a \gramor \bar{a} \gramor \tau}		& \textrm{\textbf{prefixes}} \\[2pt]

	\multicolumn{2}{l}{P,Q ::= }	& \textrm{\textbf{processes}} \\[2pt]
			& \alpha.P				& \textrm{prefix} \\[2pt]
	\gramor & P \paral Q			& \textrm{parallel} \\[2pt]
	\gramor & P + Q					& \textrm{nondeterministic choice} \\[2pt]
	\gramor & \smallsum{i}p_iP_i	& \textrm{internal probabilistic choice} \\[2pt]
	\gramor & (\nu a)P				& \textrm{restriction} \\[2pt]
	\gramor & !P					& \textrm{replication} \\[2pt]
	\gramor & 0						& \textrm{nil}
\end{array}
\]
\begin{figure}[tbp]
	$
	\begin{array}{ll@{\textrm{\hspace{15pt}}}ll}
		\textrm{ACT} &
			\bigfrac
				{}
				{\alpha.P \labarr{\alpha} \delta(P)} &

		\textrm{RES} &
			\bigfrac
				{P  \labarr{\alpha} \mu \qquad \alpha \neq a,\outp{a}}
				{(\nu a)P \labarr{\alpha} (\nu a) \mu } \\[20pt]

		\textrm{SUM1} &
			\bigfrac
				{P \labarr{\alpha} \mu }
				{P + Q \labarr{\alpha} \mu} &

		\textsc{PAR1} &
			\bigfrac
				{P \labarr{\alpha} \mu }
				{P \paral Q  \labarr{\alpha} \mu \paral Q} \\[20pt]

		\textrm{COM} & 
			\bigfrac
				{P  \labarr{a} \delta(P') \quad Q \labarr{\outp{a}}{} \delta(Q')}
				{P \paral Q \labarr{\tau} \delta(P' \paral Q')} &
	
		\textrm{PROB} &
			\bigfrac
				{}
				{\smallsum{i} p_iP_i \labarr{\tau} \smallsum{i}[p_i]\delta(P_i)} \\[20pt]

		\textrm{BANG1} &
			\bigfrac
				{P  \labarr{\alpha} \mu }
				{!P  \labarr{\alpha} \mu \paral !P} &

		\textrm{BANG2} & 
			\bigfrac
				{P  \labarr{a} \delta(P_1) \quad P \labarr{\outp{a}}{} \delta(P_2)}
				{!P \labarr{\tau} \delta(P_1 \paral P_2 \paral !P)}
	\end{array}
	$
	\caption{The semantics of \ccsp{}. SUM1 and PAR1 have corresponding
right rules SUM2 and PAR2, omitted for simplicity.}
	\label{fig:ccsp_sem}
\end{figure}
We will also use the notation $P_1 +_p P_2$ to represent a binary sum
$\smallsum{i} p_iP_i$ with $p_1=p$ and $p_2=1-p$.

The semantics of a \ccsp{} term is a probabilistic automaton defined inductively
on the basis of the syntax according to the rules in Figure
\ref{fig:ccsp_sem}.  We write $s \labarr{a} \mu$ when
$(s,a,\mu)$ is a transition of the probabilistic automaton.
We also denote by $\mu\paral Q$ the measure
$\mu'$ such that $\mu'(P\paral Q) = \mu(P)$ for all processes $P$
and $\mu'(R) = 0$ if $R$ is not of the form $P\paral Q$.
Similarly $(\nu a)\mu = \mu'$ such that $\mu'( (\nu a)P ) = \mu(P)$.

A transition of the form $P \labarr{a} \delta(P')$, i.e. a
transition having for target  a Dirac measure, corresponds
to a transition of a non-probabilistic automaton (a standard labeled transition
system). Thus, all the rules of \ccsp{} imitate the ones of CCS except
from PROB. The latter models the internal probabilistic choice: a
silent $\tau$ transition is available from the sum to a measure containing
all of its operands, with the corresponding probabilities. 

Note that in the produced probabilistic automaton, all transitions to
non-Dirac measures are silent. This is similar to the \emph{alternating
model} \cite{Hansson:89:SRTS}, however our case is more general because
the silent and non-silent transitions are not necessarily alternated. 
On the other hand, with respect to to the simple probabilistic
automata the fact that the probabilistic transitions are silent looks as
a restriction. However, it has been proved by Bandini and Segala
\cite{Bandini:01:ICALP} that the simple probabilistic automata and the
alternating model are essentially equivalent, so, being in the middle, 
our model is equivalent as well.

\section{A variant of CCS with explicit scheduler}
\label{sec:ccss}
In this section we present a variant of CCS in which the scheduler is explicit,
in the sense that it has a specific syntax and its behavior is defined
by the operational semantics of the calculus. We will refer to this calculus as
\ccss{}. Processes in \ccss{} contain labels that allow us to refer to a
particular sub-process. A scheduler also behaves like a process, using however
a different and much simpler syntax, and its purpose is to guide the execution
of the main process using the labels that the latter provides.
A \emph{complete process} is a process running in parallel with a
scheduler, and we will formally describe their interaction by 
defining an operational semantics for complete processes.

We will present \ccss{} in an incremental way. First we define the
basic calculus \ccss{} which is
the same as \ccsp{} with the addition of the explicit scheduler. Then
we will perform an 
extensions of this basic calculus by adding choice to the scheduler
so to achieve its full expressive power. Finally, in Section
\ref{sec:application}, we outline an extension of \ccss{} with a
second \emph{independent} scheduler, 
to the purpose of making private certain nondeterministic choices.

\subsection{Syntax}
\begin{figure}[tb]
	$
	\begin{array}{c@{\hspace{20pt}}c}
		\begin{array}[t]{@{\textrm{\hspace{10pt}}} l @{} l @{\textrm{\hspace{10pt}}} l }
			\multicolumn{2}{l}{I ::= 0\,I \gramor 1\,I \gramor \epsilon\quad}	& \textrm{\textbf{label indexes}} \\[2pt]
			\multicolumn{2}{l}{L ::=  l^{I}}								& \textrm{\textbf{labels}} \\[10pt]

			\multicolumn{2}{l}{P,Q ::= }	& \textrm{\textbf{processes}} \\[2pt]
					& L\cl\alpha.P			& \textrm{prefix} \\[2pt]
			\gramor & P \paral Q			& \textrm{parallel} \\[2pt]
			\gramor & P + Q					& \textrm{nondeterministic choice} \\[2pt]
			\gramor & l\cl\smallsum{i}p_iP_i	& \textrm{internal prob. choice} \\[2pt]
			\gramor & (\nu a)P				& \textrm{restriction} \\[2pt]
			\gramor & !P					& \textrm{replication} \\[2pt]
			\gramor & 0						& \textrm{nil}

		\end{array} &
		\begin{array}[t]{@{\textrm{\hspace{10pt}}}l@{}l@{\textrm{\hspace{10pt}}}l}
			\multicolumn{2}{l}{S,T ::=}		& \textrm{\textbf{scheduler}} \\[2pt]
					& \sigma(L).S			& \textrm{schedule single action} \\[2pt]
			\gramor & \sigma(L,L).S			& \textrm{synchronization} \\[2pt]
			\gramor & 0						& \textrm{nil} \\[10pt]

			\multicolumn{2}{l}{CP ::=  P \parallel S}	& \textrm{\textbf{complete process}}
		\end{array}
	\end{array}
	$
	\caption{The syntax of the core \ccss{}}
	\label{fig:syntax}
\end{figure}

Let $a$ range over a countable set of \emph{channel names} and $l$ over a countable set of
\emph{atomic labels}. The syntax of \ccss{}, shown in Figure
\ref{fig:syntax}, is the same as the one of \ccsp{} except for the
presence of labels. These
are used to select the subprocess which \qm{performs} a
transition. Since only the operators with an initial rule can
originate a transition, we only need to assign labels to the prefix
and to the probabilistic sum. We use labels of the form $l^s$ where $l$ is an
atomic label and the index $s$ is a finite string of $0$ and $1$, possibly
empty\footnote{For simplicity we will write $l$ for $l^\epsilon$.}. Indexes are used to
avoid multiple copies of the same label in  case of replication,
which occurs dynamically due to the the bang operator. As explained in the
semantics, each time a process is replicated we  relabel it using appropriate
indexes. 

A scheduler selects a sub-process for execution on the basis of its
label, so we use $\sigma(l).S$ to represent a scheduler that selects
the process with label $l$ and continues as $S$. 
In the case of synchronization we need to select two processes
simultaneously, hence we need a scheduler the form
$\sigma(l_1,l_2).S$. A complete process
is a process put in parallel with a scheduler, for example $l_1\cl a.l_2\cl b \parallel
\sigma(l_1).\sigma(l_2)$. Note that for processes with an infinite execution path we
need schedulers of infinite length.

\subsection{Semantics}
\begin{figure}[tbp]
	$
	\begin{array}{ll@{\textrm{\hspace{5pt}}}ll}
		\textrm{ACT} &
			\bigfrac
				{}
				{l\cl\alpha.P \parallel \sigma(l).S \labarr{\alpha} \delta(P \parallel S)} &

		\textrm{RES} &
			\bigfrac
				{P \parallel S \labarr{\alpha} \mu \quad \alpha \neq a,\outp{a}}
				{(\nu a)P \parallel S \labarr{\alpha} (\nu a)\mu} \\[22pt]

		\textrm{SUM1} &
			\bigfrac
				{P \parallel S \labarr{\alpha} \mu}
				{P + Q \parallel S \labarr{\alpha} \mu} &

		\textsc{PAR1} &
			\bigfrac
				{P \parallel S \labarr{\alpha} \mu}
				{ P \paral Q \parallel S \labarr{\alpha} \mu \paral Q } \\[22pt]

		\textrm{COM} & \multicolumn{3}{l}{
			\bigfrac
				{P \parallel \sigma(l_1) \labarr{a} \delta(P' \parallel 0) \qquad
				 Q \parallel \sigma(l_2) \labarr{\outp{a}}{} \delta(Q' \parallel 0)}
				{P \paral Q \parallel \sigma(l_1,l_2).S \labarr{\tau} \delta(P'\paral Q' \parallel S)}
		} \\[22pt]
	
		%\textrm{BANG} & \multicolumn{3}{l}{
		%	\bigfrac
		%		{\rho_0(P) \paral \rho_1(!P) \parallel S \labarr{\alpha} \mu}
		%		{!P \parallel \rho^{-1}(S) \labarr{\alpha} \mu} 
		%} \\[22pt]

		\textrm{BANG1} &
			\bigfrac
				{P \parallel S \labarr{\alpha} \mu}
				{!P \parallel S \labarr{\alpha} \rho_0(\mu) \paral \rho_1(!P)} &

		\textrm{PROB} &
			\bigfrac
				{}
				{l\cl \smallsum{i}p_iP_i \parallel \sigma(l).S
					\labarr{\tau} \smallsum{i}[p_i]\delta(P_i \parallel S) } \\[22pt]

		\textrm{BANG2} & \multicolumn{3}{l}{
			\bigfrac
				{P \parallel \sigma(l_1) \labarr{a} \delta(P_1 \parallel 0) \qquad
				 P \parallel \sigma(l_2) \labarr{\outp{a}}{} \delta(P_2 \parallel 0)}
				{!P \parallel \sigma(l_1,l_2).S \labarr{\tau}
					\delta(\rho_0(P_1)\paral \rho_{10}(P_2)\paral \rho_{11}(!P) \parallel S)}
		}
	\end{array}
	$
	\caption{The semantics of \ccss{}. SUM1 and PAR1 have corresponding
right rules SUM2 and PAR2, omitted for simplicity.}
	\label{fig:ccss_sem}
\end{figure}

The operational semantics of the \ccss{}-calculus is given in terms of
probabilistic automata defined inductively
on the basis of the syntax, according to the rules shown in Figure \ref{fig:ccss_sem}.

ACT is the basic communication rule.
In order for $l\cl\alpha.P$ to perform $\alpha$, the scheduler should select this
process for execution, so the scheduler needs to be of the form
$\sigma(l).S$. After the execution the complete process
will continue as $P \parallel S$. The RES rule models restriction on
channel $a$: communication on this channel is not allowed by the restricted
process.
Similarly to the section \ref{sec:ccsp}, we denote by $(\nu a)\mu$ the measure
$\mu'$ such that $\mu'( (\nu a)P \parallel S) = \mu(P \parallel S)$ for
all processes $P$ and $\mu'(R\parallel S) = 0$ if $R$ is not of the form $(\nu a)P$.
SUM1 models nondeterministic choice. If $P \parallel S$ can perform a
transition to $\mu$, which means that $S$ selects one of the labels of $P$,
then $P+Q \parallel S$ will perform the same transition, i.e. the
branch $P$ of the choice will be selected and $Q$ will be discarded. For example
\[ l_1\cl a.P + l_2\cl b.Q \parallel \sigma(l_1).S \labarr{a} \delta(P \parallel S) \]
Note that the operands
of the sum do not have labels, the labels belong to the subprocesses of $P$ and $Q$.
In the case of nested choices, the scheduler must go deep and select the label
of a prefix, thus resolving all the choices at once.

PAR1 has a similar behavior for parallel composition. The scheduler selects
$P$ to perform a transition on the basis of the label. The difference
is that in this case $Q$ is not discarded; it remains in the
continuation. $\mu \paral Q$ denotes the measure
$\mu'$ such that $\mu'(P\paral Q \parallel S) = \mu(P \parallel S)$.
%The side condition means that $\mu'(R\parallel S)$ is
%$0$ everywhere except when $R$ is of the form $P'\paral Q$ for some $P'$, in
%which case it coincides with $\mu(P'\parallel S)$.
COM models synchronization.
If $P \parallel \sigma(l_1)$ can perform the action $a$ and $Q \parallel \sigma(l_2)$
can perform $\bar{a}$, then $\sigma(l_1, l_2)$, scheduling both $l_1$ and $l_2$ at
the same time, can synchronize the two. PROB models internal probabilistic choice.
Note that the scheduler cannot affect the outcome of the choice, it can only
schedule the choice as a whole (that's why a probabilistic sum has a label)
and the process will move to a measure containing all the operands with
corresponding probabilities.

Finally, BANG1 and BANG2 model replication. The rules are the same as in \ccsp{}, with the
addition of a re-labeling operator $\rho_k$. The reason for this is
that we want to avoid ending up with multiple copies of the same label as the
result of replication, since this would create ambiguities in scheduling as
explained in section \ref{sec:detlabeling}.
$\rho_k(P)$ replaces all labels $l^s$ inside $P$ with $l^{sk}$, and it is defined as
\begin{eqnarray*}
	\rho_k(l^s\cl \alpha.P) &=& l^{sk} \cl \alpha.\rho_k(P)
\end{eqnarray*}
and homomorphically on the other operators (for instance $\rho_k(P\paral Q) =
\rho_k(P) \paral $ $\rho_k(Q)$).
We also denote by $\rho_k(\mu)$ the measure $\mu'$ such that
$\mu'(\rho_k(P) \parallel S) = \mu(P \parallel S)$.
Note that we relabel only the resulting process, not the continuation of the scheduler:
there is no need for relabeling the scheduler since we are free to choose the continuation
as we please.

Let us give an example of how BANG1 and relabeling work.
Let $P =$ \mbox{$l_1\cl a.l_2\cl b$}. To prove a transition for $!P \parallel \sigma(l_1).S$
we have to prove it for $P$ and then relabel the resulting process:
\begin{prooftree}
	\AxiomC{}
	\LeftLabel{ACT}
    \UnaryInfC{
		$l_1\cl a.l_2\cl b \parallel \sigma(l_1).S \labarr{a}
			\delta(l_2\cl b \parallel S)$
	}
	\LeftLabel{BANG1}
    \UnaryInfC{
		$!(l_1\cl a.l_2\cl b) \parallel \sigma(l_1).S \labarr{a}
			\delta(l_2^0\cl b \paral !(l_1^1\cl a.l_2^1\cl b) \parallel S)$
	}
\end{prooftree}
As we can see in the example, when a process $!\, P$ is
activated, the spawned copy of $P$ is relabeled
by adding $0$ to the index of all the labels, and $!P$ is relabeled by adding $1$.
So the labels of $\rho_0(P)$ and $\rho_1(!P)$ will be disjoint.
As remarked above, the continuation $S$ is not relabeled, if we want to perform
$b$ after $a$ then $S$ should start with $\sigma(l_2^0)$.

\subsection{Deterministic labelings}
\label{sec:detlabeling}
The idea in \ccss{} is that a \emph{syntactic} scheduler will be able to completely
solve the nondeterminism of the process, without needing to rely on a
\emph{semantic} scheduler at the level of the automaton. This means that
the execution of a process in parallel with a scheduler should be
fully probabilistic.
To achieve this we will impose a condition on the labels that we can use in \ccss{}
processes. A \emph{labeling} is an assignment of labels to the prefixes and
probabilistic sums of a
process. We will require all labelings to be \emph{deterministic} in the 
following sense.

\begin{definition}
	A labeling of a process $P$ is \emph{deterministic} iff for all schedulers
	$S$ there is only one transition rule $P\parallel S \labarr{\alpha} \mu$
	that can be applied and the labelings of all processes $P'$ such that
   	$\mu(P')>0$	are also deterministic.
\end{definition}
A labeling is \emph{linear} iff all labels are pairwise disjoint. We can show
that linear labelings are preserved by transitions, which leads to the following
proposition.
\begin{proposition}
	A linear labeling is deterministic.
\end{proposition}

There are labelings that are deterministic without being linear. In fact, such
labelings will be the means by which we hide information from the scheduler.
However, the property of being deterministic is crucial since it implies that
the scheduler will resolve all the nondeterminism of the process.
\begin{proposition}
Let $P$ be a \ccss{} process with a deterministic labeling. Then for all
schedulers $S$, the automaton produced by $P\parallel S$ is fully probabilistic.
\end{proposition}

\section{Expressiveness of the syntactic scheduler}\label{sec:expressiveness}
\ccss{} with deterministic labelings allows us to separate probabilities from
nondeterminism in a straightforward way: a process in parallel with a scheduler
behaves in a fully probabilistic way and the nondeterminism arises from the fact
that we can have many different schedulers. We may now ask the question: how powerful
are the syntactic schedulers wrt the semantic ones, i.e. those defined
directly over the automaton?

Let $P$ be as \ccsp{} process and $P_\sigma$ be the \ccss{} process obtained
from $P$ by applying a linear labeling. We say that the semantic scheduler $\zeta$
of $P$ is equivalent to the syntactic scheduler $S$ of $P_\sigma$,
written $\zeta \sim_P S$, iff the automata\footnote{Note that
with a slight abuse of notation we will use a process to denote its corresponding
probabilistic automaton.}
$etree(P,\zeta)$ and $P_\sigma \parallel S$ are probabilistically
bisimilar in the sense of \cite{Segala:95:NJC}. 
%Cat: I think the following is redundant 
%We remind that $etree(P,\zeta)$ is a fully probabilistic
%automaton whose states are executions of $P$, and $P_\sigma\parallel S$ is the fully probabilistic
%automaton generated by the semantics of \ccss{}.

A scheduler $S$ is \emph{non-blocking} for a process $P$ if it always schedules some
transitions, except when $P$ itself is blocked. Since 
semantic schedulers are usually not allowed to block, we will restraint ourselves to non-blocking syntactic
schedulers to obtain a $1-1$ correspondence. Let $Sem(P)$ be the set
of the semantic schedulers
for the process $P$, and $Syn(P_\sigma)$ be the set of the non-blocking syntactic schedulers for
process $P_\sigma$. The following result holds for pure CCS processes (that is, \ccsp{}
processes without probabilistic choice).
\begin{proposition}
	Let $P$ be a pure CCS process and let $P_\sigma$ be a 
	\ccss{} process obtained by adding a linear labeling to $P$. Then
	\[
	\begin{array}{l}
		\forall \zeta \in Sem(P) \ \exists S \in Syn(P_\sigma) : \zeta \sim_P S
			\quad\textrm{and} \\
		\forall S \in Syn(P_\sigma) \ \exists \zeta \in Sem(P) : \zeta \sim_P S
	\end{array}
	\]
	\label{prop:equiv_scheduler_1}
\end{proposition}

\subsection{The scheduler in the presence of probabilistic choice}
In Proposition
\ref{prop:equiv_scheduler_1} we considered pure CCS processes
in which the execution tree has only
one possible execution. Now consider a process $P$ containing
an internal probabilistic choice. Even if we fix the scheduler, the
outcome of the choice is not always the same, so $P \parallel S$
could produce different executions. As a consequence, the syntactic
schedulers we have defined are not enough to give us back all
the semantic ones. Consider the process
$ P  = l\cl (l_1\cl a +_p l_2\cl b) $.
After the probabilistic choice, either $l_1\cl a$ or $l_2\cl b$ will
be available, but we cannot know which one. As a consequence, we cannot
create a scheduler that selects $a$ or $b$, whatever is available. In
fact, it's not even possible to create a non-blocking scheduler at all,
both $\sigma(l).\sigma(l_1)$ and $\sigma(l).\sigma(l_2)$ will block on some
executions.

The problem here is that the process can make choices that are independent
from the scheduler, so the latter should adapt its behavior
to the outcome of these choices. To achieve this, we extend \ccss{}
by adding a \emph{scheduler choice} construct. The new syntax and semantics are
displayed in Figure \ref{fig:schedulerchoice}. A scheduler can be
the sum $\sum_iS_i$ of several schedulers, and the outcome of the
probabilistic choice (in the process) will determine the one to
activate. In the case where more than one could
be activated at the same time, we give preference to the first one in
the sum, so the scheduler
still behaves in a deterministic way.

\begin{figure}[tbp]
	\centering
	$
	\begin{array}{l@{\hspace{30pt}}l}
		S ::= \ldots \gramor \smallsum{i}S_i
		&
		\begin{array}{ll}
			\textrm{TEST} &
				\bigfrac
					{P \parallel S_i \labarr{\alpha} \mu \qquad
						\forall j < i: P \parallel S_j \nrightarrow  }
					{P \parallel\smallsum{i}S_i \labarr{\alpha} \mu }
		\end{array}
	\end{array}
	$
	\caption{Adding scheduler choice to \ccss}
	\label{fig:schedulerchoice}
\end{figure}

In our previous example, we can use the scheduler $\sigma(l)(\sigma(l_1) +
\sigma(l_2))$ which will produce $a$ or $b$ depending on the outcome of the
probabilistic choice. With the scheduler choice, we can retrieve the full power
of the semantic scheduler for full \ccsp{} processes.
\begin{proposition}
	Proposition \ref{prop:equiv_scheduler_1} holds for full \ccsp{} processes
	if we extend schedulers with scheduler choice.
	\label{prop:equiv_scheduler_2}
\end{proposition}

\subsection{Using non-linear labelings}
Up to now we are using only linear labelings which, as we saw, give us the
whole power of semantic schedulers. However, we can construct non-linear
labelings that are still deterministic, that is there is still only one
transition possible at any time even though we have multiple occurrences of
the same label. There are various cases of useful non-linear labelings.

\begin{proposition}
	Let $P$,$Q$ be \ccss{} processes with deterministic labelings (not
	necessarily disjoint). The following labelings are all deterministic:
	\begin{eqnarray}
		& l\cl (P +_p Q)		        \label{eq:detlab1} \\
		& l_1\cl a.P + l_2\cl b.Q		\label{eq:detlab2} \\
		& (\nu a)(\nu b)(l_1\cl a. P + l_1\cl b.Q \paral l_2\cl \bar{a}) \label{eq:detlab3}
	\end{eqnarray}
	\vspace{-15pt}
	\label{prop:detlab}
\end{proposition}
Consider the case where $P$ and $Q$ in the above proposition share the same labels.
In (\ref{eq:detlab1}) the scheduler cannot select an action inside $P,Q$, it must select
the choice itself. After the choice, only one of $P,Q$ will be available so there will be
no ambiguity in selecting transitions. The
case (\ref{eq:detlab2}) is similar but with nondeterministic choice. Now the guarding
prefixes must have different labels, since the scheduler should be able to resolve the
choice, however after the choice only one of $P,Q$ will be available. Hence, again, the multiple
copies of the labels do not constitute a problem. In (\ref{eq:detlab3}) we allow the same label
on the guarding prefixes of a nondeterministic choice. This is because the guarding
channels $a,b$ are restricted and only one of the corresponding output actions is
available ($\bar{a}$). As a consequence, there is no ambiguity in selecting
transitions. A scheduler $\sigma(l_1,l_2)$ can only perform a synchronization on
$a$, even though $l_1$ appears twice.

However, using multiple copies of a label limits the power of the
scheduler, since the labels provide information about the outcome of a
probabilistic choice (and allow the scheduler to choose different 
strategies through the use of the scheduler choice). In fact, this is
exactly the technique we will use to archive the goals described in Section
\ref{sec:introduction}. Consider for example the process:
\begin{equation}
	l\cl(l_1\cl \bar{a} +_p l_1\cl \bar{a}) \paral l_2\cl a.P \paral l_3\cl a.Q
	\label{eq:example1}
\end{equation}
From Proposition \ref{prop:detlab}(\ref{eq:detlab1}) this labeling is deterministic. However, since
both branches of the probabilistic sum have the same label $l_1$,
the scheduler cannot resolve the choice between $P$ and $Q$ based on the outcome of
the choice. There is still nondeterminism: the scheduler $\sigma(l).\sigma(l_1,l_2)$
will select $P$ and the scheduler $\sigma(l).\sigma(l_1,l_3)$ will select $Q$. However
this selection will be independent from the outcome of the probabilistic choice.

Note that we did not impose any direct restrictions on the schedulers, we still consider
all possible syntactic schedulers for the process (\ref{eq:example1}) above. However,
having the same label twice limits the power of the syntactic
schedulers with respect to the
semantic ones. This approach has the advantage that the restrictions are limited to
the choices with the same label. We already know that having pairwise different labels
gives the full power of the semantic scheduler. So the restriction is local to the place
where we, intentionally, put the same labels.

\section{Testing relations for \ccss{} processes}\label{sec:testing}
Testing relations \cite{DeNicola:84:TCS} are a method of comparing processes by considering
their interaction with the environment. A \emph{test} is a process running in parallel with
the one being tested and which can perform a distinguished action $\omega$ that represents
success. Two processes are testing equivalent if they can pass the same tests.
This idea is very useful for the analysis of security protocols, as suggested in
\cite{Abadi:99:IC}, since a test can be seen as an adversary who interferes with a communication
agent and declares $\omega$ if an attack is successful. Then two processes are testing
equivalent if they are vulnerable to the same attacks.

In the probabilistic setting we take the approach of \cite{Jonsson:2001:HANDBOOK} which
considers the exact probability of passing a test (in contrast to
\cite{Palamidessi:05:TCS} which considers only the ability to pass a test with
probability non-zero (may-testing) or one (must-testing)). This approach leads to
the definition of two preorders $\may$ and $\must$. $P\may Q$ means that the
if $P$ can pass $O$ then $Q$ can also pass $O$ with the same probability. 
$P \must Q$ means that if $P$ always passes $O$ with at least some probability then
$Q$ always passes $O$ with at least the same probability.

A labeling of a process is \emph{fresh} (with respect to a set $\cal
P$ of processes) if it is linear and its labels do not appear
in any other process in $\cal P$. A test $O$ is a \ccss{} process with a fresh labeling, containing
the distinguished action $\omega$. Let $\it{Test}_{\cal P}$ denote the set of all
tests with respect to $\cal P$ and let $(\nu)P$ denote the restriction on all channels of
$P$, thus allowing only $\tau$ actions. 
We define $\pom(P, S, O)$ to be the probability
of the set of executions of the fully probabilistic automaton $(\nu)(P \paral O) \parallel S$
that contain $\omega$. Note that this set can be produced as a countable union of disjoint
cones so its probability is well-defined.
\begin{definition}
	Let $P,Q$ be \ccss{} processes. We define must and may testing preorders as follows:
	\[\begin{array}{lcl}
		P \may Q & \textrm{ iff } &
			\forall O \in \it{Test}_{P,Q} \ \forall S_P \in Syn((\nu)(P\paral O)) \ \exists S_Q \in Syn((\nu)(Q\paral O)): \\[2pt]
		&&	\qquad \pom(P,S_P,O) \le \pom(Q,S_Q,O) \\[4pt]
		P \must Q & \textrm{ iff } &
			\forall O \in \it{Test}_{P,Q} \ \forall S_Q \in Syn((\nu)(Q\paral O)) \ \exists S_P \in Syn((\nu)(P\paral O)): \\[2pt]
		&&	\qquad \pom(P,S_P,O) \le \pom(Q,S_Q,O)
	\end{array}\]
\end{definition}
We also define $\mayeq,\musteq$ to be the equivalences induced by $\may,\must$
respectively.

A context $C$ is a process with a hole. A preorder $\weaker$ is a precongruence if
$P \weaker Q$ implies $C[P] \weaker C[Q]$ for all contexts $C$. May and must testing
are precongruences if we restrict to contexts with fresh labelings and
without occurrences of $+$. This result is essentially an adaptation
to our framework of the analogous precongruence property in
\cite{Yi:92:IFIP}.
\begin{proposition}
	Let $P,Q$ be \ccss{} processes such that $P\may Q$ and let $C$ be a context with
	a fresh labeling and in which $+$ does not occur. Then $C[P] \may C[Q]$. Similarly for $\must$.
	\label{prop:precongruence}
\end{proposition}
This also implies that $\mayeq,\musteq$ are congruences.
Note that $P,Q$ in the above proposition are not required to have linear labelings,
$P$ might include multiple occurrences of the same label thus limiting the power of
the schedulers $S_P$. This shows the locality of the scheduler's restriction:
some choices inside $P$ are hidden from the scheduler but the rest of the
context is fully visible.

If we remove the freshness condition then Proposition
\ref{prop:precongruence} is no longer true. Let $P = l_1\cl a.l_2\cl b$, $Q = l_3\cl
a.l_4\cl b$
and $C = l\cl(l_1\cl a.l_2\cl c +_p [\,])$.
We have $P \mayeq Q$ but $C[P],C[Q]$ can be separated by the test
$O = \bar{a}.\bar{b}.\omega \paral \bar{a}.\bar{c}.\omega$ (The labeling
is omitted for simplicity since tests always have fresh labelings.)
It is easy to see that $C[Q]$ can pass the test with probability $1$ by
selecting the correct branch of $O$ based on the outcome of the probabilistic
choice. In $C[P]$ this is not possible because of the labels $l_1,l_2$ that
are common in $P,C$.

We can now state the result that we announced in Section \ref{sec:introduction}.
%Let $P,Q$ be \ccss{} processes and $C$ a context without bang and
%with a fresh labeling.
%Now consider the process $R = l\cl(C[l_1\cl \tau.P] +_p C[l_1\tau.Q])$. The labels of $C$
%are duplicated in each branch of the probabilistic choice (remember that
%this is allowed by Proposition \ref{prop:detlab}). So a scheduler $S$ controlling
%this process will not be able to see the outcome of the choice and it will
%have to control both $C[\tau.P]$ and $C[\tau.Q]$ in the same way, until it reaches
%$\tau.P$ (or $\tau.Q$) itself. Then consider the process
%$R' = C[l\cl(P +_p Q)]$. We can create a similar scheduler $S'$ who immitates $S$
%and performs the choice when $S$ reaches $P$ or $Q$. As a consequence $R,R'$
%will be indistinguishable by the environment.
\begin{theorem}
	Let $P,Q$ be \ccss{} processes and $C$ a context
	with a fresh labeling and  without
        occurrences of bang. Then
%\footnote{For simplicity we omit the label $l$ of $+_p$ and the label $l_1$ of $\tau.P$ and $\tau.Q$.}
	\begin{eqnarray*}
		l\cl (C[l_1\cl\tau.P] +_p C[l_1\cl\tau.Q])	& \mayeq & C[l\cl(P +_p Q)] \quad \textrm{and} \\
		l\cl (C[l_1\cl\tau.P] +_p C[l_1\cl\tau.Q])	& \musteq & C[l\cl(P +_p Q)]
	\end{eqnarray*}
	\vspace{-20pt}
	\label{thm:distoversum}
\end{theorem}
The proof is given in the appendix.

There are two crucial points in the above Theorem. The first is that the labels
of the context are copied, thus the scheduler cannot distinguish between $C[l_1\cl\tau.P]$ and
$C[l_1\cl\tau.Q]$ based on the labels of the context. The second is that $P,Q$ are protected by a $\tau$ action labeled by
the same label $l_1$. This is to ensure that in the case of a nondeterministic
sum ($C = R + []$) the scheduler cannot find out whether the second operand of
the choice is $P$ or $Q$ unless it commits to selecting the second operand.
For example let $R = l_1\cl(l_2\cl a +_{0.5} 0), P = l_3\cl a, Q = 0$.
Then $R_1 = (R + a) +_{0.1} (R + 0)$ is not testing equivalent
to  $R_2 = R + (a +_{0.1} 0)$  since they can be separated by
$O = \outp{a}.\omega$ and a scheduler that resolves $R+a$ to $a$ and $R+0$ to $R$.
However, if we take $R_1' = (R + l\cl \tau.a) +_{0.1} (R + l\cl\tau. 0)$ then
$R_1'$ is testing equivalent to $R_2$ since the scheduler will have to resolve
both branches of $R_1'$ in the same way (even though we still have non-determinism).

The problem with replication is simply the persistence of the processes.
It is clear that $!P +_p !Q$ cannot be equivalent in any way to $!(P +_p Q)$
since the first replicates only one of $P,Q$ while the second replicates both.
However Theorem \ref{thm:distoversum} together with Proposition \ref{prop:precongruence}
imply that
\begin{equation}
	C'[l\cl(C[l_1\cl\tau.P] +_p C[l_1\cl\tau.Q])] \mayeq C'[C[l\cl(P +_p Q)]]
	\label{eq:twocontexts}
\end{equation}
where $C$ is a context without bang and $C'$ is a context without $+$. The same is also true for
$\musteq$. This means that we can lift the sum towards the root of the context until we reach
a bang. Intuitively we cannot move the sum outside the bang since each replicated copy
must perform a different probabilistic choice with a possibly different outcome.

Theorem \ref{thm:distoversum} shows that the probabilistic choice is indeed private to
the process and invisible to the scheduler. The process can perform it at any time, even
in the very beginning of the execution, without making any difference to an outside
observer.

\section{An application to security}\label{sec:application}
In this section we discuss an application of our framework to anonymity. In particular, we show how to 
specify the Dining Cryptographers protocol \cite{Chaum:88:JC} so that it is robust to scheduler-based attacks.
We first propose a method to encode \emph{secret value passing}, which will turn out to be useful
for the specification:
\begin{eqnarray}
	l\cl c(x).P & \defeq & \smallsum{i} l\cl cv_i.P[v_i/x] \\
	l\cl \bar{c}\ang{v}.P & \defeq & l\cl\outp{cv}.P
\end{eqnarray}
This is the usual encoding of value passing in CSS except that
we use the same label in all the branches of the nondeterministic
sum. To ensure that the resulting labeling will be deterministic we
should restrict the channels $cv_i$ and make sure that there will
be at most one output on $c$. We will write $(\nu c)P$ for
$(\nu cv_1)\ldots(\nu cv_n)P$. For example, the labeling of the following
process is deterministic:
\[ 
	(\nu c)(l_1\cl c(x).P \paral l\cl(l_2\cl\bar{c}\ang{v_1} +_p l_2\cl\bar{c}\ang{v_2})) \qquad
\]
This case is a combination of the cases (\ref{eq:detlab1}) and (\ref{eq:detlab3}) of
Proposition \ref{prop:detlab}. The two outputs on $c$ are on different branches of
the probabilistic sum, so during an execution at most one of them will be available.
Thus there is no ambiguity in scheduling the sum produced by $c(x)$. The scheduler
$\sigma(l).\sigma(l_1,l_2)$ will perform a synchronization on $cv_1$ or $cv_2$,
whatever is available after the probabilistic choice. In other words, using the
labels we manage to hide the information about which value was transmitted to $P$.

\subsection{Dining cryptographers with probabilistic master}\label{sec:dc}
\begin{figure}[tbp]
\begin{eqnarray*}
	Master &\defeq& l_1\cl \smallsumb{i=0}{2} p_i
		(\underbrace{\bar{m}_0\ang{i\eqeq 0}}_{l_2} \paral
		 \underbrace{\bar{m}_1\ang{i\eqeq 1}}_{l_3} \paral
		 \underbrace{\bar{m}_2\ang{i\eqeq 2}}_{l_4}) \\[2pt]
	Crypt_i &\defeq&
		\underbrace{ m_i(pay) }_{l_{5,i}}.
		\underbrace{ c_{i,i}(coin_1) }_{l_{6,i}}.
		\underbrace{ c_{i,i\oplus 1}(coin_2) }_{l_{7,i}}.
		\underbrace{ \outp{out}_i\ang{pay\otimes coin_1\otimes coin_2} }_{l_{8,i}}	\\[2pt]
	Coin_i & \defeq& l_{9,i}\cl(
		(\underbrace{ \bar{c}_{i,i}\ang{0} }_{l_{10,i}} \paral
		 \underbrace{ \bar{c}_{i\ominus 1,i}\ang{0} }_{l_{11,i}}) +_{0.5}
		(\underbrace{ \bar{c}_{i,i}\ang{1} }_{l_{10,i}} \paral
		 \underbrace{ \bar{c}_{i\ominus 1,i}\ang{1} }_{l_{11,i}} )
		) \\[2pt]
	Prot & \defeq & (\nu \vec{m})(
		Master \paral (\nu \vec{c})(
			\smallprodb{i=0}{2} Crypt_i \paral
			\smallprodb{i=0}{2} Coin_i ))
\end{eqnarray*}
	\caption{Encoding of the dining cryptographers with probabilistic master}
	\label{fig:dining}
\end{figure}

The problem of the Dining Cryptographers is the following: Three cryptographers are dining together. At the end of the
dinner, the bill has to be paid by either one of them or by another agent called the
master. The master decides who will pay and then informs each of them separately
 whether he has to pay or not. The cryptographers
would like to find out whether the payer is the master or one of them.
However, in the latter case, they also wish to keep the payer anonymous.  

The Dining Cryptographers Protocol (DCP) solves the above problem as follows: each
cryptographer tosses a fair coin which is visible to himself and his neighbor to the
right. Each cryptographer checks the two adjacent coins and, if he is 
not paying, announces
{\it agree} if they are the same and {\it disagree} otherwise. However, the
paying cryptographer will say the opposite. It can be proved that if the number
of {\it disagrees} is even, then the master is paying; otherwise, one of the
cryptographers is paying \cite{Chaum:88:JC}.

An external observer $O$ is supposed to see only the three announcements $\outp{out}_i\ang{\ldots}$. 
As discussed in \cite{Bhargava:05:CONCUR}, DCP satisfies anonymity if 
we abstract from their order. If their order is observable, on the contrary, then
a scheduler can reveal the identity of the payer to $O$ 
simply by forcing the payer to make his announcement first. Of course, 
this is possible only if the scheduler is unrestricted and can choose its 
strategy depending on the decision of the master (or on the results of the coins). 

In our framework we can solve the problem by giving a specification of the DCP in which the choices 
of the master and of the coins are made invisible to the scheduler. 
The specification is shown in Figure \ref{fig:dining}. We use some meta-syntax for brevity:
The symbols $\oplus$ and $\ominus$ represent
the addition and subtraction modulo 3, while $\otimes$ represents the addition modulo 2 (xor).
The notation $i \eqeq n$ stands for $1$ if $i=n$ and $0$ otherwise.

There are many sources of nondeterminism: the
order of communication between the master and the cryptographers,
the order of reception of the coins, and the order of the announcements. 
The crucial points of our specification, which  make the nondeterministic choices independent 
from the probabilistic ones, are:  (a) all communications internal to the 
protocol (master-cryptographers and cryptographers-coins) are done by secret value passing, 
and (b) in each probabilistic choice the different branches have the same labels. For example,
all branches of the master contain an output on $m_0$, always labeled by
$l_2$, but with different values each time. 

Thanks to the above independence, the specification satisfy strong probabilistic anonymity. There are various equivalent 
definitions of this property, we follow here the version presented in \cite{Bhargava:05:CONCUR}. Let
$\vec{o}$ represent an observable (the sequence of announcements), and $p_S(\vec{o} \paral \bar{m}_i\ang{1})$ represent
the conditional probability, under the scheduler $S$, that the protocol produces $\vec{o}$ given that the 
master has selected Cryptographer $i$ as the payer. 

\begin{proposition}[Strong probabilistic anonymity]
The protocol in Figure \ref{fig:dining} satisfies the following property: for all schedulers $S$ and all observables
$\vec{o}$,
$p_S(\vec{o} \paral \bar{m}_0\ang{1}) =
 p_S(\vec{o} \paral \bar{m}_1\ang{1}) =
 p_S(\vec{o} \paral \bar{m}_2\ang{1})$
\end{proposition}
Note that different schedulers will produce different traces (we still have nondeterminism) but
they will not depend on the choice of the master.

Some previous treatment of the DCP, including
\cite{Bhargava:05:CONCUR}, had solved the problem of the
leak of information due to too-powerful schedulers by simply
considering as observable sets of announcements instead than
sequences. Thus one could think that using a true concurrent
semantics, for instance event structures, would solve the problem.  
We would like to remark that this is false: true concurrency 
would weaken the scheduler  enough in the case of the
DCP, but not in general. For instance, it would not help in the
anonymity example in the introduction. 

\subsection{Dining cryptographers with nondeterministic master}
\begin{figure}[tbp]
	\[
	\begin{array}{l@{\hspace{30pt}}l}
		\begin{array}{ll}
			P & ::= \ldots \gramor l\cl \{P\} \\[2pt]
			CP & ::= P \parallel S, T
		\end{array}
		&
		\begin{array}{ll}
			\textrm{INDEP} &
			\bigfrac
				{P \parallel T \labarr{\alpha} \mu }
				{\begin{array}{c}
					l\cl \{P\} \parallel \sigma(l).S,T \labarr{\alpha} \mu' \\
					\textrm{where }	\mu'(P' \parallel S,T') = \mu(P' \parallel T')
				\end{array}}
		\end{array}
	\end{array}
	\]
	\caption{Adding an \qm{independent} scheduler to the calculus}
	\label{fig:independent}
\end{figure}

We sketch here a method to hide also certain nondeterministic choices from the scheduler, 
and we show an application to the variant of the Dining Cryptographers with nondeterministic master. 

First we need to extend the calculus with the concept of a second \emph{independent} scheduler $T$ that
we assume to solve the nondeterministic choices that we want to make transparent to the 
main scheduler $S$. The new syntax and semantics
are shown in Figure \ref{fig:independent}. $l:\{P\}$ represents a process where
the scheduling of $P$ is protected from the main scheduler $S$. The scheduler $S$ can
\qm{ask} $T$ to schedule $P$ by selecting the label $l$. Then $T$ resolves
the nondeterminism of $P$ as expressed by the INDEP rule. Note that we need to
adjust also the other rules of the semantics to take $T$ into account, but
this change is straightforward. % (maybe put it in the appendix?)
We assume that $T$ does not collaborate with $S$ so we do not need to
worry about the labels in $P$. 

To model the dining cryptographers with nondeterministic
master we replace the $Master$ process in Figure \ref{fig:dining} by the
following one.
\[
Master \defeq l_1\cl \big\{ \smallsumb{i=0}{2} l_{12,i}\cl\tau.
(\underbrace{\bar{m}_0\ang{i\eqeq 0}}_{l_2} \paral
		 \underbrace{\bar{m}_1\ang{i\eqeq 1}}_{l_3} \paral
		 \underbrace{\bar{m}_2\ang{i\eqeq 2}}_{l_4}) \big\}
\]
Essentially we have replaced the probabilistic choice by a \emph{protected} nondeterministic one.
Note that the labels of the operands are different but this is not a problem
since this choice will be scheduled by $T$. Note also that after the choice we
still have the same labels $l_2,l_3,l_4$, however the labeling is still
deterministic, similarly to the case \ref{eq:detlab2} of Proposition
\ref{prop:detlab}.

In case of a nondeterministic selection of the culprit, and a probabilistic anonymity protocol, the 
notion of strong probabilistic anonymity has not been established yet, although some 
possible definitions have been discussed in \cite{Bhargava:05:CONCUR}. 
Our framework makes it possible to give a natural and precise definition.

\begin{definition}[Strong probabilistic anonymity for nondeterministic selection of the culprit]
A protocol with nondeterministic selection of the culprit satisfies strong probabilistic anonymity iff 
for all observables $\vec{o}$, schedulers $S$, and independent
schedulers $T_1,T_2$ which select different culprits, we have:
$p_{S,T_1}(\vec{o}) = p_{S,T_2}(\vec{o})$. 
\end{definition}

We can prove the above property for our protocol:

\begin{proposition}
The DCP with nondeterministic selection of the culprit specified in this section 
satisfies strong probabilistic anonymity.
\end{proposition}

%The key here is that
%we compare $T_1$ and $T_2$ using the same scheduler $S$, since we assume that
%$S$ independent from $T$ thus it cannot depend on the latter.

\section{Conclusion and Future work}\label{sec:conclusion}
We have proposed a process-calculus approach to the problem of limiting the 
power of the scheduler so that it does not reveal the outcome of hidden random choices, and we have shown 
its applications to the specification of information-hiding protocols. We have also discussed a feature, 
namely the distributivity  of certain contexts over random choices,  
that makes our calculus appealing for verification. Finally, we have considered the probabilistic 
testing preorders and shown that  they are precongruences in our calculus. 

Our plans for future work are in two directions: (a) we would like 
 to investigate the possibility of giving a game-theoretic characterization of our notion of scheduler, 
and (b) we would like to incorporate our ideas in some existing probabilistic model checker, for instance PRISM. 

\subsubsection*{Acknowledgments.} We would like to thank Vincent Danos 
for having pointed out to us an attack to 
the Dining Cryptographers protocol based on the order of the
scheduler, which has inspired this work. 

\bibliographystyle{splncs}

\begin{thebibliography}{10}

\bibitem{Vardi:85:FOCS}
Vardi, M.Y.:
\newblock Automatic verification of probabilistic concurrent finite-state
  programs.
\newblock In: Proc. of the 26th Annual Symp. on Foundations of
  Computer Science, IEEE Computer Society Press (1985)
  327--338

\bibitem{Hansson:89:SRTS}
Hansson, H., Jonsson, B.:
\newblock A framework for reasoning about time and reliability.
\newblock In: Proc. of the 10th Symposium on Real-Time Systems,
  IEEE Computer Society Press (1989)  102--111

\bibitem{Yi:92:IFIP}
Yi, W., Larsen, K.G.:
\newblock Testing probabilistic and nondeterministic processes.
\newblock In: Proc. of the 12th IFIP International Symposium on Protocol
  Specification, Testing and Verification, North Holland (1992)

\bibitem{Segala:95:PhD}
Segala, R.:
\newblock Modeling and Verification of Randomized Distributed Real-Time
  Systems.
\newblock PhD thesis, Department of Electrical Engineering and Computer
  Science, Mass\-a\-chu\-setts Insti\-tute of Tech\-no\-logy (1995) Available
  as Technical Report MIT/LCS/TR-676.

\bibitem{Segala:95:NJC}
Segala, R., Lynch, N.:
\newblock Probabilistic simulations for probabilistic processes.
\newblock Nordic Journal of Computing \textbf{2} (1995)

\bibitem{Hansson:90:SRTS}
Hansson, H., Jonsson, B.:
\newblock A calculus for communicating systems with time and probabitilies.
\newblock In: Proc. of the 11th Symposium on Real-Time Systems,  
IEEE Computer Society Press (1990)  278--287

\bibitem{Bandini:01:ICALP}
Bandini, E., Segala, R.:
\newblock Axiomatizations for probabilistic bisimulation.
\newblock In: Proc. of the 28th International Colloquium on Automata,
  Languages and Programming. LNCS 2076,
  Springer (2001)  370--381

\bibitem{Andova:02:PhD}
Andova, S.:
\newblock Probabilistic process algebra.
\newblock PhD thesis, Technische Universiteit Eindhoven (2002)

\bibitem{Mislove:03:EXPRESS}
Mislove, M., Ouaknine, J., Worrell, J.:
\newblock Axioms for Probability and Nondeterminism.
\newblock In: Proc.\ of the 10th Int.\ Wksh.\
  on Expressiveness in Concurrency (EXPRESS '03). Volume~96 of ENTCS, Elsevier (2004)

\bibitem{Palamidessi:05:TCS}
Palamidessi, C., Herescu, O.M.:
\newblock A randomized encoding of the $\pi$-calculus with mixed choice.
\newblock Theoretical Computer Science \textbf{335} (2005)  373--404\\
  \url{http://www.lix.polytechnique.fr/~catuscia/papers/prob\_enc/report.%
pdf}.

\bibitem{Deng:05:BookJW}
Deng, Y., Palamidessi, C., Pang, J.:
\newblock Compositional reasoning for probabilistic finite-state behaviors.
\newblock In: Processes, Terms and Cycles: Steps on the Road to Infinity.
  LNCS 3838.
\newblock Springer (2005)  309--337\\
  \url{http://www.lix.polytechnique.fr/~catuscia/papers/Yuxin/BookJW/par.%
pdf}.

\bibitem{Sokolova:04:BookChapt}
Sokolova, A., Vink, E.d.:
\newblock Probabilistic automata: system types, parallel composition and
  comparison.
\newblock In: Validation of Stochastic Systems: A Guide to Current
  Research. LNCS 
  2925.
\newblock Springer (2004)  1--43

\bibitem{Jonsson:2001:HANDBOOK}
Jonsson, B., Larsen, K.G., Yi, W.:
\newblock Probabilistic extensions of process algebras.
\newblock In: Handbook of Process
  Algebras.
\newblock Elsevier (2001)  685--710

\bibitem{Chatzikokolakis:05:TCS}
Chatzikokolakis, K., Palamidessi, C.:
\newblock A framework for analyzing probabilistic protocols and its application
  to the partial secrets exchange.
\newblock Theoretical Computer Science. To appear. A short version of
  this paper appeared in the {\em Proc. of the Symp. on Trustworthy
  Global Computing}, LNCS 3705,  146-162. Springer, 2005.\\
  \url{http://www.lix.polytechnique.fr/~catuscia/papers/PartialSecrets/TC%
Sreport.pdf}.

\bibitem{deAlfaro:01:CONCUR}
de~Alfaro, L., Henzinger, T.A., Jhala, R.:
\newblock Compositional methods for probabilistic systems.
\newblock In: Proceedings of CONCUR 2001. LNCS 2154, Springer (2001)

\bibitem{Mitchell:06:TCS}
Mitchell, J.C., Ramanathan, A., Scedrov, A., Teague, V.:
\newblock A probabilistic polynomial-time process calculus for the analysis of
  cryptographic protocols.
\newblock Theoretical Computer Science \textbf{353} (2006)  118--164

\bibitem{Canetti:06:WODES}
Canetti, R., Cheung, L., Kaynar, D., Liskov, M., Lynch, N., Pereira, O.,
  Segala, R.:
\newblock Task-structured probabilistic i/o automata.
\newblock In: Proc. of the 8th Int. Workshop on Discrete Event
  Systems (WODES'06),  (2006)

\bibitem{Canetti:06:DISC}
Canetti, R., Cheung, L., Kaynar, D.K., Liskov, M., Lynch, N.A., Pereira, O.,
  Segala, R.:
\newblock Time-bounded task-{PIOA}s: {A} framework for analyzing security
  protocols.
\newblock In: Proc. of DISC '06. LNCS 4167, Springer (2006)  238--253

%\bibitem{Canetti:07:Rep}
%Canetti, R., Cheung, L., Lynch, N., Pereira, O.:
%\newblock On the role of scheduling in simulation-based security.
%\newblock Cryptology ePrint Archive, Report 2007/102 (2007)

\bibitem{DeNicola:84:TCS}
Nicola, R.D., Hennessy, M.C.B.:
\newblock Testing equivalences for processes.
\newblock Theoretical Computer Science \textbf{34} (1984)  83--133

\bibitem{Abadi:99:IC}
Abadi, M., Gordon, A.D.:
\newblock A calculus for cryptographic protocols: The spi calculus.
\newblock Information and Computation \textbf{148} (1999)  1--70

\bibitem{Chaum:88:JC}
Chaum, D.:
\newblock The dining cryptographers problem: Unconditional sender and recipient
  untraceability.
\newblock Journal of Cryptology \textbf{1} (1988)  65--75

\bibitem{Bhargava:05:CONCUR}
Bhargava, M., Palamidessi, C.:
\newblock Probabilistic anonymity.
\newblock In: Proc. of CONCUR 2005. LNCS 3653, Springer (2005)  171--185\\
  \url{http://www.lix.polytechnique.fr/~catuscia/papers/Anonymity/concur.%
pdf}.

\end{thebibliography}

\appendix
\section{Proofs}
In this appendix we give the proof of the main technical result of our paper.
\vskip 5mm \noindent
	\textbf{Theorem \ref{thm:distoversum}}
	Let $P,Q$ be \ccss{} processes and $C$ a context
	with a fresh labeling and  without
        occurrences of bang. Then
	\begin{eqnarray*}
		l\cl( C[l_0\cl\tau.P] +_p C[l_0\cl\tau.Q] )	& \mayeq & C[l\cl(P +_p Q)] \quad \textrm{and} \\
		l\cl( C[l_0\cl\tau.P] +_p C[l_0\cl\tau.Q] )	& \musteq & C[l\cl(P +_p Q)]
	\end{eqnarray*}
	\vspace{-20pt}
\begin{proof}\ \\
Since we will always use the label $l$ for all probabilistic sum $+_p$, and
$l_0$ for $\tau.P$ and $\tau.Q$, we will omit these labels to make
the proof more readable.
We will also denote $(1-p)$ by $\bar{p}$.

Let $R_1 = C[\tau.P] +_p C[\tau.Q]$ and $R_2 = C[P +_p Q]$.
We will prove that for all tests $O$ and for all schedulers
$S_1 \in Syn((\nu)(R_1\paral O))$ there exists $S_2 \in Syn((\nu)(R_2\paral O))$ such
that $\pom(R_1, S_1, O) = \pom(R_2,S_2,O)$ and vice
versa. This implies both $R_1 \mayeq R_2$ and $R_1 \musteq R_2$.

Without loss of generality we assume
that tests do not perform internal actions, but only synchronizations with the tested
process. First, it is easy to see that
\begin{eqnarray}
	\pom(P +_p Q, \sigma(l).S, O) &=&
		p\ \pom(P, S, O) + \bar{p}\ \pom(Q, S, O) \label{eq:pom1}\\
	\pom(l_1\cl a.P, \sigma(l_1,l_2).S, O) & = & \pom(P, S, O')  \label{eq:pom2}
\end{eqnarray}
where
$(\nu)(l_1\cl a.P\paral O)\parallel \sigma(l_1,l_2).S \labarr{\tau}
	\delta((\nu)(P\paral O' \parallel S))$.

In order for the scheduler of $R_1$ to be non-blocking, it has to be of the form
$\sigma(l).S_1$, since the only possible transition of $R_1$ is the probabilistic
choice labeled by $l$. By (\ref{eq:pom1}) we have
\[
	\pom(C[\tau.P] + C[\tau.Q], \sigma(l).S_1, O) =
		p\ \pom(C[\tau.P],S_1,O) + \bar{p}\ \pom(C[\tau.Q],S_1,O)
\]
The proof will be by induction on the structure of $C$. Let $O$ range over tests with
fresh labelings, let $S_1$ range over nonblocking schedulers for both $C[\tau.P]$ and $C[\tau.Q]$ (such
that $\sigma(l).S_1$ is a nonblocking scheduler for $R_1$) and let $S_2$ range over
nonblocking schedulers for $R_2$. The induction hypothesis is:
\[
\begin{array}{l}
	\Rightarrow)\  \forall O\ \forall S_1\ \exists S_2 : \\
		\qquad p\ \pom(C[\tau.P],S_1,O) + \bar{p}\ \pom(C[\tau.Q],S_1,O) = \pom(C[P +_p Q], S_2, O)
		\quad \textrm{and} \\
	\Leftarrow)\ \forall O\ \forall S_2\ \exists S_1: \\
		\qquad p\ \pom(C[\tau.P],S_1,O) + \bar{p}\ \pom(C[\tau.Q],S_1,O) = \pom(C[P +_p Q], S_2, O)
\end{array}
\]
We have the following cases for $C$:
\begin{itemize}
	\item Case $C = []$. Trivial.
	\item Case $C = l_1\cl a.C'$\\
	The scheduler $S_1$ of $C[\tau.P]$ and $C[\tau.Q]$ has to be of the form
	$S_1 = \sigma(l_1, l_2).S_1'$ where $l_2$ is the label of a $\outp{a}$
	prefix in $O$ (if no such prefix exists then the case is trivial).

	A scheduler of the form $\sigma(l_1,l_2).S$ can schedule any process of the form $l_1\cl a.X$
	(with label $l_1$) giving the transition:
	\[ (\nu)(l_1\cl a.X \paral O) \parallel \sigma(l_1,l_2).S \labarr{\tau}
		\delta((\nu)(X \paral O') \parallel S)
	\]
	and producing always the same $O'$.
	The probability $\pom$ for these processes will be given by equation (\ref{eq:pom2}).

	Thus for ($\Rightarrow$) we have
	\[
	\begin{array}{r @{\textrm{\hspace{5pt}}} cl @{\textrm{\hspace{5pt}}} l}
		\multicolumn{3}{l}{
			p\ \pom(l_1\cl a.C[\tau.P],\sigma(l_1,l_2).S_1',O) +
			\bar{p}\ \pom(l_1\cl a.C[\tau.Q],\sigma(l_1,l_2).S_1',O) } \\[3pt]
		&=&	p\ \pom(C'[\tau.P], S_1', O') + \bar{p}\ \pom(C'[\tau.Q], S_1', O')	& \textrm{(\ref{eq:pom2})} \\[3pt]
		&=& \pom(C'[P +_p Q], S_2', O')						& \textrm{Ind. Hyp.} \\[3pt]
		&=& \pom(l_1\cl a.C'[P +_p Q], \sigma(l_1,l_2).S_2', O)	& \textrm{(\ref{eq:pom2})} \\[3pt]
		&=& \pom(R_2, S_2, O)
	\end{array}
	\]
	For ($\Leftarrow$) we can perform the above derivation in the opposite direction, given
	that a scheduler for $R_2 = l_1\cl a.C'[P +_p Q]$ must be of the form $S_2 = \sigma(l_1,l_2).S_2'$.

	\item Case $C = C' \paral R$\\
	Since we only consider contexts with
	fresh labelings, $R \paral O$ is itself a test, and
	\begin{equation}
		\pom(X\paral R, S, O) = \pom(X, S, R\paral O)	\label{eq3}
	\end{equation}
	Thus for ($\Rightarrow$) we have
	\[
	\begin{array}{r @{\textrm{\hspace{5pt}}} cl @{\textrm{\hspace{5pt}}} l}
		\multicolumn{3}{l}{
			p\ \pom(C'[\tau.P]\paral R, S_1, O) +
			\bar{p}\ \pom(C'[\tau.Q] \paral R, S_1, O) }		\\[3pt]

		&=& p\ \pom(C'[\tau.P], S_1, R \paral O) +
			\bar{p}\ \pom(C'[\tau.Q], S_1, R \paral O)			& \textrm{(\ref{eq3})} \\[3pt]

		&=& \pom(C'[P +_p Q], S_2, R\paral O)					& \textrm{Ind. Hyp.} \\[3pt]
		&=& \pom(C'[P +_p Q]\paral R, S_2, O)					& \textrm{(\ref{eq3})} \\[3pt]
		&=& \pom(R_2, S_2, O)
	\end{array}
	\]
	For ($\Leftarrow$) we can perform the above derivation in the opposite direction.
	\vspace{5pt}

	\item Case $C = l_1\cl(C' +_q R)$ \\
	Since we consider only contexts with fresh labelings, the labels of $C'$ are
	disjoint from those of $R$, thus the scheduler of a
	process of the form $l_1\cl(C'[X] +_q R)$ must be of the form
	$S = \sigma(l_1).(S_C + S_R)$ where $S_C$ is a scheduler containing labels
	of $C'[X]$ and $S_R$ a scheduler containing labels of $R$. Moreover
	\begin{eqnarray}
		\lefteqn{\pom(l_1\cl(C'[X] +_q R), S, O)} \nonumber\\
		&=&	q\ \pom(C'[X], S_C+S_R, O) + \bar{q}\ \pom(R, S_C+S_R, O) \nonumber\\
		&=&	q\ \pom(C'[X], S_C, O) + \bar{q}\ \pom(R, S_R, O) \label{eq:pom3}
	\end{eqnarray}
	As a consequence, the scheduler $S_1$ of $C[\tau.P]$ and $C[\tau.Q]$ has to be of the form
	$S_1 = \sigma(l_1).(S_C + S_R)$.

	For ($\Rightarrow$) we have
	\[
	\begin{array}{r @{\textrm{\hspace{5pt}}} cl @{\textrm{\hspace{5pt}}} l}
		\multicolumn{4}{l}{
			p\ \pom(l_1\cl( C'[\tau.P] +_q R ), S_1, O) +
			\bar{p}\ \pom(l_1\cl( C'[\tau.Q] +_q R ), S_1, O) }		\\[3pt]

		&=& q(p\ \pom(C'[\tau.P], S_C, O) + \bar{p}\ \pom(C'[\tau.Q], S_C, O)) + \\
		&&	\bar{q}\ \pom(R, S_R, O)		& \textrm{(\ref{eq:pom3})} \\

		&=& q\ \pom(C'[P+_pQ]), S_C', O) + \\
		&&	\bar{q}\ \pom(R, S_R, O)			& \textrm{Ind. Hyp.} \\

		&=& \pom(l_1\cl( C'[P+_pQ] +_q R ), \sigma(l_1).(S_C'+S_R), O) & \textrm{(\ref{eq:pom3})} \\
		&=& \pom(R_2, S_2, O)
	\end{array}
	\]
	For ($\Leftarrow$) we can perform the above derivation in the opposite direction, given
	that a scheduler for $R_2 = l_1\cl(C'[P +_p Q] +_q R)$ must be of the form
	$S_2 = \sigma(l_1).(S_C' + S_R)$.
	\vspace{5pt}

	\item Case $C = C' + R$\\
	Consider the process $C'[l_0\cl\tau.P] + R$. The scheduler $S_1$ of this process has to
	choose between $C'[l_0\cl\tau.P]$ and $R$.

	There are two cases to have a transition using the SUM1, SUM2 rules.
	\begin{itemize}
		\item[i)] Either
		$ (\nu)(C'[l_0\cl\tau.P] + R \paral O) \parallel S_R \labarr{\alpha} \mu $
		such that $(\nu)( R \paral O) \parallel S_R \labarr{\alpha} \mu$. In this
		case
		\begin{equation}
			\pom(C'[l_0\cl\tau.P] + R, S_R, O) = \pom(R, S_R, O) \label{eq4}
		\end{equation}

		\item[ii)] Or
		$ (\nu)(C'[l_0\cl\tau.P] + R \paral O) \parallel S_C \labarr{\alpha} \mu $
		such that $(\nu)(C'[l_0\cl\tau.P] \paral O) \parallel S_C \labarr{\alpha} \mu$.
		In this case
		\begin{equation}
			\pom(C'[l_0\cl\tau.P] + R, S_C, O) = \pom(C'[l_0\cl\tau.P], S_C, O) \label{eq5}
		\end{equation}
	\end{itemize}

	Now consider the process $C'[l_0\cl\tau.Q] + R$. Since $P$ and $Q$ are behind
	the $l_1\cl\tau$ action, this process has exactly the same visible labels
	as $C'[l_0\cl\tau.P] + R$. Thus $S_R$ and $S_C$ will select
	$R$ and $C'[l_0\cl\tau.Q]$ respectively and the equations (\ref{eq4}) and (\ref{eq5})
	will hold.

	In the case (i) ($S = S_R$) we have:
	\[
	\begin{array}{r @{\textrm{\hspace{5pt}}} cl @{\textrm{\hspace{5pt}}} l}
		\multicolumn{4}{l}{
			p\ \pom(C'[\tau.P] + R , S_R, O) +
			\bar{p}\ \pom(C'[\tau.Q] + R), S_R, O) }		\\[3pt]

		&=&	p\ \pom(R , S_R, O) +
			\bar{p}\ \pom(R, S_R, O) 		& \textrm{(\ref{eq4})} \\[3pt]

		&=&	\pom(R , S_R, O) 				\\[3pt]
		&=&	\pom(C'[P +_p Q] + R , S_R, O) 	\\[3pt]

		&=& \pom(R_2, S_2, O)
	\end{array}
	\]

	In the case (ii) ($S = S_C$) we have:
	\[
	\begin{array}{r @{\textrm{\hspace{5pt}}} cl @{\textrm{\hspace{5pt}}} l}
		\multicolumn{4}{l}{
			p\ \pom(C'[\tau.P] + R , S_C, O) +
			\bar{p}\ \pom(C'[\tau.Q] + R), S_C, O) }		\\[3pt]

		&=&	p\ \pom(C'[\tau.P] , S_C, O) +
			\bar{p}\ \pom(C'[\tau.Q], S_C, O) 		& \textrm{(\ref{eq5})} \\[3pt]

		&=&	\pom(C'[P +_p Q], S_C', O)  		& \textrm{Ind. Hyp.} \\[3pt]

		&=&	\pom(C'[P +_p Q] + R , S_C', O) 	\\[3pt]

		&=& \pom(R_2, S_2, O)
	\end{array}
	\]
	For ($\Leftarrow$) we can perform the above derivation in the opposite direction.
	\vspace{5pt}

	\item Case $C = (\nu a)C'$\\
	The process $(\nu)( (\nu a)C'[X] \paral O )$ has the same transitions as
	$(\nu)(C'[X] \paral (\nu a)O)$. The result follows by the induction hypothesis.

%	\vspace{100pt}
%	We have $R_1 = l\cl((C'[P] + R) +_p  (C'[Q]+R))$ and $R_2 = C'[l\cl(P+_pQ)]+ R$.
%	PROBLEM \\
%	Counter example $R = (a +_{0.1} 0), P = a, Q = 0, p = 0.5, O = \outp{a}.\omega$. \\
%	$R + (a +_{0.1} 0)$ is not equivalent to $(R + a) +_{0.1} (R + 0)$,
%	in the first the best is to select $R$ since it has 0.5 chance of producing $a$.
%	In the second we can wait to see the outcome of the probabilistic choice, if
%	$R+a$ is chosen then we select $a$, if $R+0$ is chosen then we select
%	$R$. This gives probability $0.55$ of passing the test.

\end{itemize}
$\hfill\Box$
\end{proof}

\end{document}